\documentclass[aps,prb,twocolumn,groupedaddress,nofootinbib]{revtex4-1}
\usepackage{amsmath,bm}
\usepackage{graphicx}
\usepackage[bottom]{footmisc}
\usepackage{float}

\begin{document}
\title{Density Wave States in the Presence of an External Magnetic Field}


\author{Ian E. Powell and Sudip Chakravarty}
\affiliation{Mani L Bhaumik Institute for Theoretical Physics\\Department of Physics and Astronomy, University of California Los Angeles, Los Angeles, California 90095, USA}
\date{\today}

\begin{abstract}

We investigate the effect that density-wave states have on the Hofstadter Butterfly.  We first review the problem of the $d$-density wave on a square lattice and then numerically solve the $d$-density wave problem when an external magnetic field is introduced.  As the $d$-density wave condensation strength is tuned the spectrum evolves through three topologically distinct butterflies, and an unusual quantum Hall effect is observed.  The chiral $p+ip$-density wave state demonstrates drastically different Hofstadter physics--inducing a destruction of the gaps in the butterfly which causes electrons' cyclotron orbits to not obey any type of Landau quantization, and the creation of a large gap in the spectrum with Hall conductance $\sigma_{xy}$=0.  To investigate the quantum phases in the system we perform a multifractal analysis of the single particle wavefunctions.  We find that tuning the $d$-density wave strength at a generic value of magnetic flux controls a metal-metal transition at charge neutrality where the wavefunction multifractality occurs near band touching events.  In the $p+ip$ case we observe another metal-metal transition near a band touching event which is seperated by a quasi-insulating island state occuring at charge neutrality near strip dimerization of the lattice.
\end{abstract}


\maketitle
\section{Preliminaries}

When electrons in two dimensions are subjected to a periodic potential of a crystalline lattice and a uniform magnetic field, the two competing length scales, that of the Landau levels and the crystalline lattice result in a quantum fractal spectra. Despite the beauty and the complexity of the structure it has received very little attention because length scales are typically severely mismatched. It is only recently the Hofstadter butterfly spectra in bilayer graphene has opened up the possibility of emergent behavior within a fractal landscape~\cite{Kim}. In this paper we wish to take the further step and ask what if the lattice exhibits further  symmetry breaking, in particular density waves. This may lead to novel emergent behavior, especially exotic superconducting states.

Unlike Cooper pair condensation (particle-particle condensation) density wave states are comprised of particle-hole condensates.  The particle-hole condensate wavefunction does not have to obey the same spin/orbital antisymmetry requirements that Cooper pair wavefunctions do because particles and holes are distinct objects.  A particularly interesting density wave state is the $d_{x^2-y^2}$-density wave, also known as the staggered flux state.  The staggered flux state is visualized as a series of staggered currents on the bonds of the square lattice\cite{Chetan}.  We briefly review particle-hole condensation in this angular momentum channel on the square lattice in the following.

On the mean field level the single particle Hamiltonian for electrons in an external magnetic field with singlet particle-hole pairing in the $d_{x^2-y^2}$ channel on the square lattice in position space is written as
\cite{ChakravartyXunGoswami, Hofstadter} 
\begin{equation} \label{eq:square} 
\begin{split}
H=\sum_{n,m }\left(-t_1 + i\frac{W_0}{4}(-1)^{n+m}\right)e^{i \phi_x}| m+1, n \rangle \langle m, n | \\
+\left(-t_2 - i\frac{W_0}{4}(-1)^{n+m}\right)e^{i \phi_y}| m, n+1
 \rangle \langle m, n | \\
 -t_3 e^{i \phi_{xy}}| m+1, n+1 \rangle \langle m, n |\\
  -t_4 e^{i \phi_{yx}}| m+1, n-1 \rangle \langle m, n | + \text{H.C.}
\end{split}
\end{equation}
where each $\phi$ is the Peierls phase associated with each unique hopping element, we have subtracted off the chemical potential, and we have included only nearest neighbor (NN) and next-nearest neighbor (NNN) terms.  For the remainder of the paper we take $t_1=t_2=t$, and omit spin indices.

When there is no external magnetic field present the staggered flux causes the unit cell's size to double--comprised of an $n+m=\text{even}$,  $n+m=\text{odd}$.  Ignoring NNN hopping we write the Hamiltonian in the absence of external magnetic field as
\begin{equation}
H=-\tilde{t}\sum_{n,m }e^{-2i \alpha_{nm}}| m+1, n \rangle \langle m, n |\\
+| m, n+1
 \rangle \langle m, n |\\
 + \text{H.C.},
\end{equation}
where we have a new variable, $\tilde{t}=\sqrt{t^2+(W_0/4)^2}$, and $\alpha_{nm}$ = arctan$(W_0/4t)(-1)^{n+m}$.  In this language the dispersion is written as
\begin{equation}
E=\pm 2 \tilde{t}\sqrt{\text{cos}^2(k_x)+\text{cos}^2(k_y)+2\text{cos}(2\alpha)\text{cos}(k_x)\text{cos}(k_y)},
\end{equation}
where $\alpha = |\alpha_{nm}|$ and 
\begin{equation}
\text{cos}(2\alpha)=\frac{1-(W_0/4t)^2}{1+(W_0/4t)^2}.
\end{equation}
We see that as the density wave strength is tuned on from 0 the dispersion evolves smoothly from the free electron case to the staggered Fermion case at $\alpha=\pi/4$.  With this in mind we rewrite the Hamiltonian as
\begin{equation}
\begin{split}
H=-\tilde{t}\sum_{n,m }e^{-2i \alpha_{nm}}| m+1, n \rangle \langle m, n |+\\
(\text{cos}^2(2\alpha)+\text{sin}^2(2\alpha))| m, n+1\rangle \langle m, n |+ \text{H.C.}, 
\end{split}
\end{equation}
which is equivalent to 
\begin{equation}
H=\text{cos}(2\alpha)H_{tb}+\text{sin}(2\alpha)H_{sf},
\end{equation}
where
\begin{equation}
H_{tb}=-\tilde{t}\sum_{n,m}| m+1, n \rangle \langle m, n |+\text{cos}(2\alpha)| m, n+1
 \rangle \langle m, n |\\
 + \text{H.C.},
\end{equation}
and
\begin{equation}
\begin{split}
H_{sf}=-\tilde{t}\sum_{n,m}-i (-1)^{m+n}| m+1, n \rangle \langle m, n |+\\
\text{sin}(2\alpha)| m, n+1
 \rangle \langle m, n |
+ \text{H.C.}.
\end{split}
\end{equation}

\section{Butterflies} 
\subsection{Nearest Neighbors}
Turning on an external magnetic field in the d-density wave problem amounts to the usual Peierls substitution \cite{Peierls}.  Taking the Landau gauge $\vec{A}=(-By,0,0)$, the $m$ direction hopping elements in the Hamiltonian (Eq. 6) are modified via $|m+1,n\rangle \rightarrow e^{-i2 \pi n \Phi/\Phi_0}|m+1,n\rangle$, where $2\pi\Phi/\Phi_0$ is the dimensionless magnetic flux penetrating an elementary plaquette.  We numerically diagonalize the Hamiltonian on a 20$\times$20 lattice and plot the energy (in units of $t$) versus $\Phi/\Phi_0$ at the highest symmetry in Fig.s 1,2 and 3. 

\begin{figure}[ht]
		 \centering \includegraphics[width=\columnwidth]{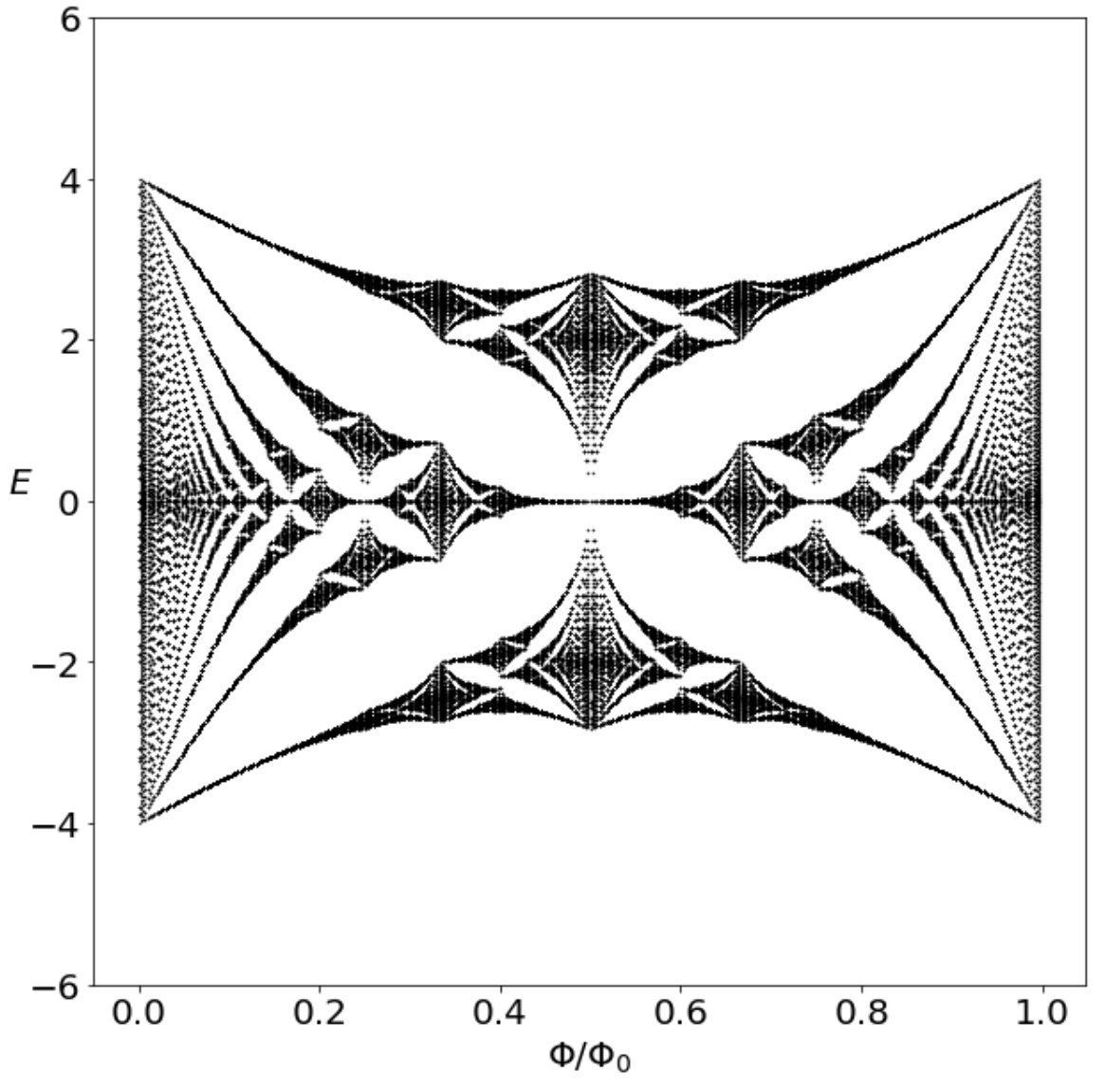}

        \caption{
                \label{fig:samplesetup} 
                Plot of the butterfly for $\alpha=0$. 
        }
\end{figure}

\begin{figure}[ht]

        \centering \includegraphics[width=\columnwidth]{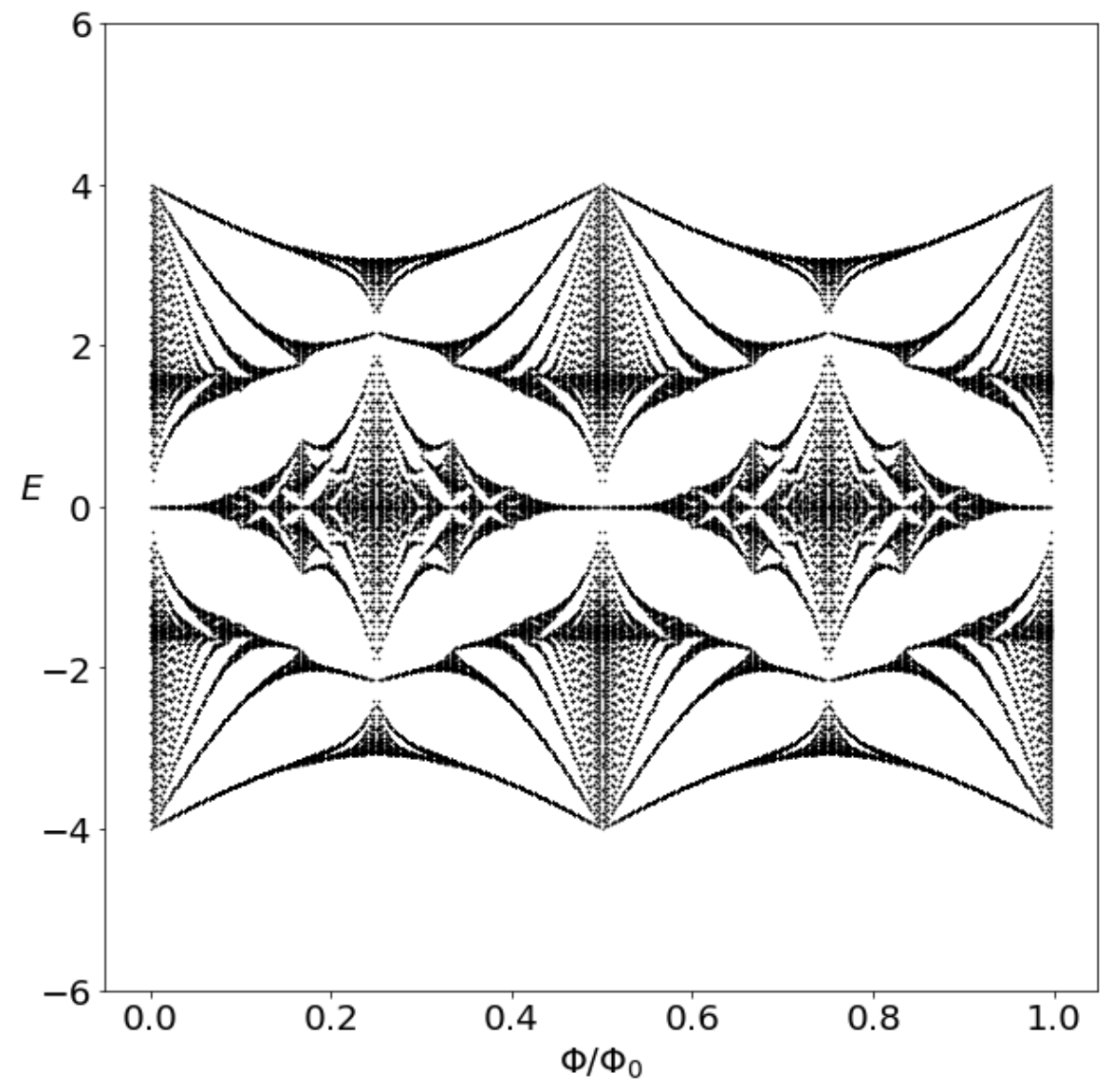}

        \caption{
                \label{fig:samplesetup} 
                Plot of the butterfly for $\alpha=\pi/8$.
        }
\end{figure}

\begin{figure}[ht]

        \centering \includegraphics[width=\columnwidth]{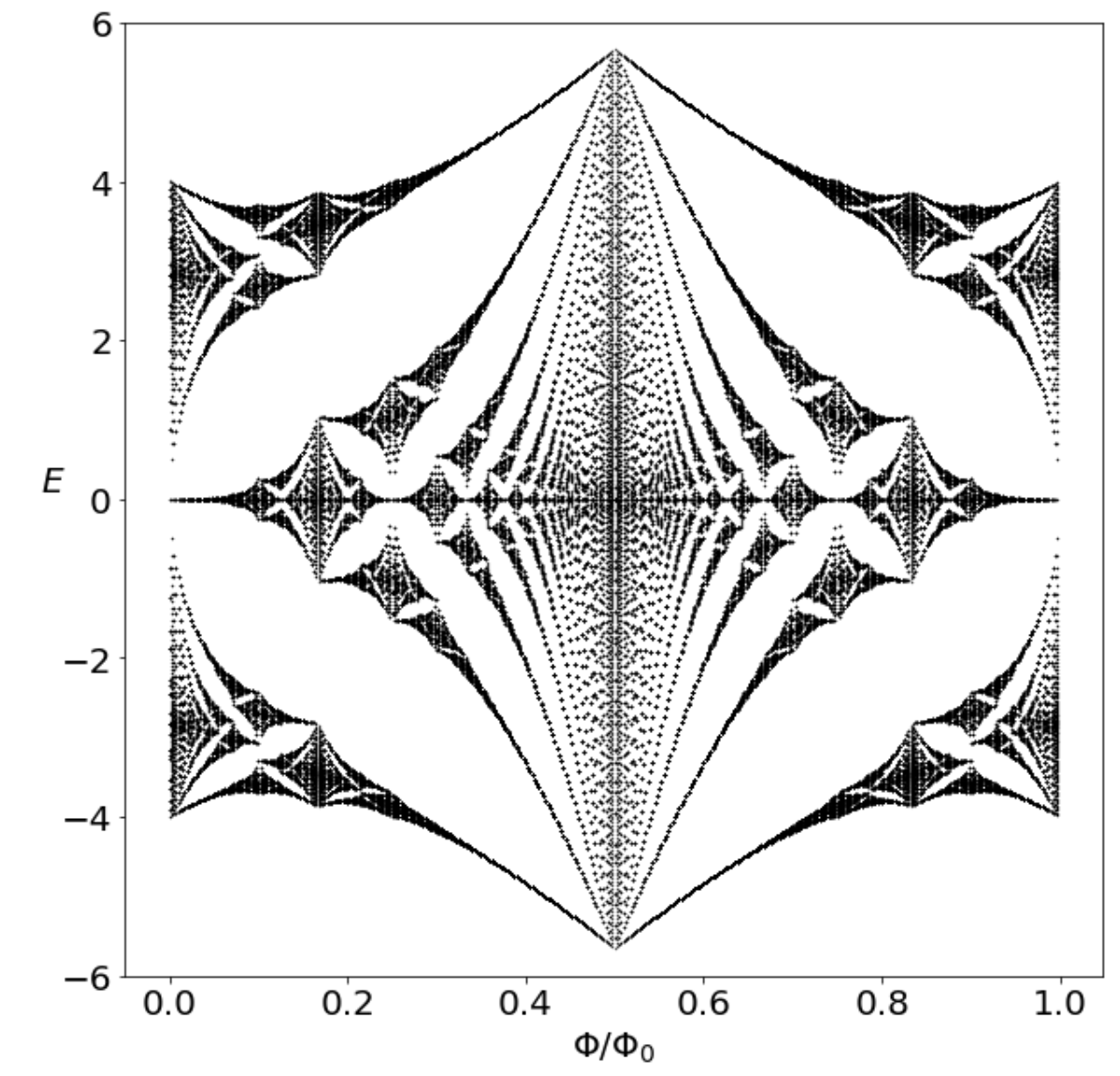}

        \caption{
                \label{fig:samplesetup} 
                Plot of the butterfly for $\alpha=\pi/4$.
        }
\end{figure}

When $\alpha=0$ we recover the usual Hofstadter butterfly, and when $\alpha=\pi/4$ we recover the ``fermionic" butterfly\cite{endrodi} governed by the form of Eq. 8.  As $\alpha$ is tuned away from 0 linear Landau levels emerge from the edges of the spectrum at $\pi$ flux, and relativistic levels emerge at 0 and 2$\pi$ flux at charge neutrality.  All emerging Landau levels are accompanied by gap openings with odd Chern number which will be discussed further in the following section.  The relativistic Landau level energy eigenvalues emerging from 0 flux are given by (see Appendix A)
\begin{equation} 
\epsilon_n = \pm 2 \sqrt{\frac{e_0 B |W_0| t}{c}n}.
\end{equation}
As $W_0$ is tuned from 0 to $4t$ the Hall conductances, $\sigma_{xy}$, change for a given flux and Fermi energy.  Due to the global nature of the transformation of the topological phase diagram (the Hofstadter butterfly) we catagorize the topologically different types of Butterflies instead of investigating topological phase transitions local to a given flux and Fermi energy in the following section.
\subsection{Topological Maps of the $d_{x^2-y^2}$-density-wave Butterfly}
To characterize the defining topological characteristics of each butterfly we start with the extremum of the transformation controlled by the density wave strength.
First of all, consider the situation when $\alpha$=$\pi/4$.  Directly from our Gauge transformed Hamiltonian we see that the total flux penetrating a plaquette is $\Phi \pm 4|\alpha|=\Phi \pm \pi$, where the plus or minus indicates that we are at an even/odd plaquette respectively.  Thus the Hamiltonian can be written as

\begin{equation}
\begin{split}
H=-\sqrt{2}t\sum_{n,m }e^{-i(\Phi+\pi)n}| m+1, n \rangle \langle m, n |\\
+| m, n+1
 \rangle \langle m, n |+ \text{H.C.},
\end{split}
\end{equation}
because the Hamiltonian in the abscence of density-wave condensation is symmetric about $\Phi=\pm \pi$.  

This observation explicitely shows that the density-wave parameter $W_0$ controls a smooth transformation between the typical butterfly and the $\pi$-shifted, or ``fermionic," butterfly.  The Hall conductances for the gaps can be written down immediately for these two extremum of the transformation via a Diophantine equation\cite{Thouless}, but because the particular Diophantine equation which governs the region $0<\alpha<\pi/4$ is not immediately obvious we follow a different prescription.

To describe the global distribution of Chern numbers in the gaps of the butterflies we closely follow work done by Naumis\cite{Naumis} on the ``Cut and Projection" solution to the Diophantine equation
\begin{equation}
\sigma_r = q\left\{\phi r +\frac{1}{2}\right\}-\frac{q}{2}.
\end{equation}
Here $\sigma$ is the Hall conductance, $r$ is the gap index, the curly braces indicate taking the fractional part of the quantity contained, and $\phi=\Phi/\Phi_0=p/q$ where $p/q$ is a fully reduced fraction.  The filling factor for a gap's Chern number at a given flux, or the ``hull function'' $f(\phi,\sigma)$, is also defined as
\begin{equation}
f(\phi,\sigma)= \{\phi\sigma_r\}.
\end{equation}
Plotting these hull functions against the flux yields what is known as a butterfly's ``skeleton."  The form of a skeleton dictates the distribution of Hall Conductances in the gaps of the butterflies.  We find the skeletons for $\alpha=0,\pi/4$ using this hull function expression (see Fig.s 4 and 5).
\begin{figure}[t]
		 \centering \includegraphics[width=\columnwidth]{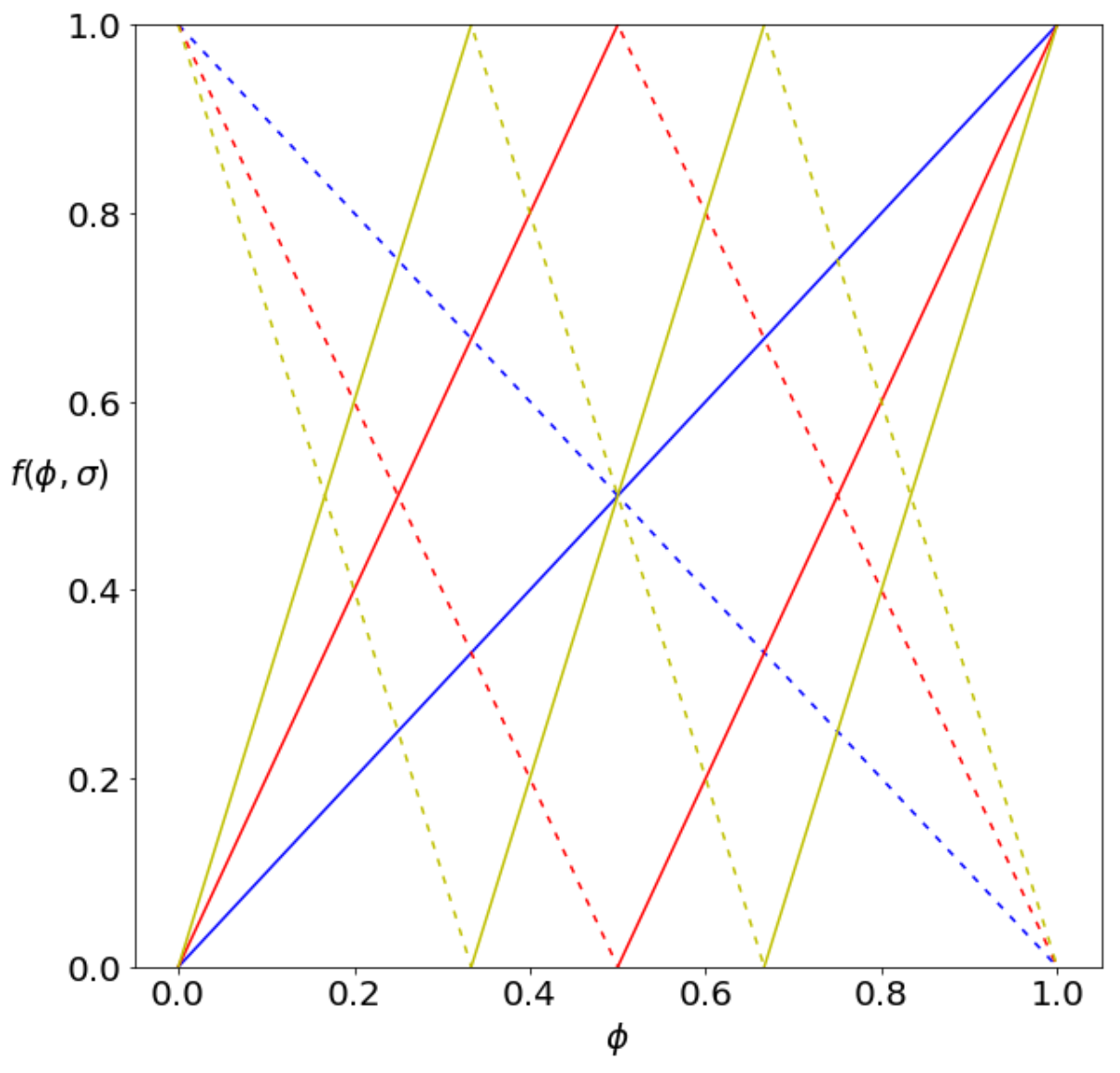}

        \caption{
                \label{fig:samplesetup} 
                Skeleton of the butterfly for $\alpha$=0.  Solid (dashed) blue lines correspond to $\sigma_{xy}$=1 (-1), solid (dashed) red lines correspond to $\sigma_{xy}$=2 (-2), and solid (dashed) yellow lines correspond to $\sigma_{xy}$=3 (-3).  
        }
\end{figure}
We choose this hull function formalism as it naturally characterizes the global structure of the different butterflies in our problem.
To construct the skeleton for values of 0$<\alpha<\pi/4$ we note the following: as soon as $W_0$ is nonzero all gaps that are not associtated with the normal butterfly, but are associated with the $\pi$ flux butterfly, emerge (in Appendix A we see that regardless of how small $W_0$ is all Landau levels indexed by $n$ emerge).  Furthermore, the Chern numbers associated with all gaps are topological invariants and thus will not change due to perturbations to the Hamiltonian.  Taking these facts into account we draw the topological map for the region 0$<\alpha<\pi/4$ as the combination of the two extremum butterfly skeletons--see Fig. 6.
\begin{figure}[H]
        \centering \includegraphics[width=\columnwidth]{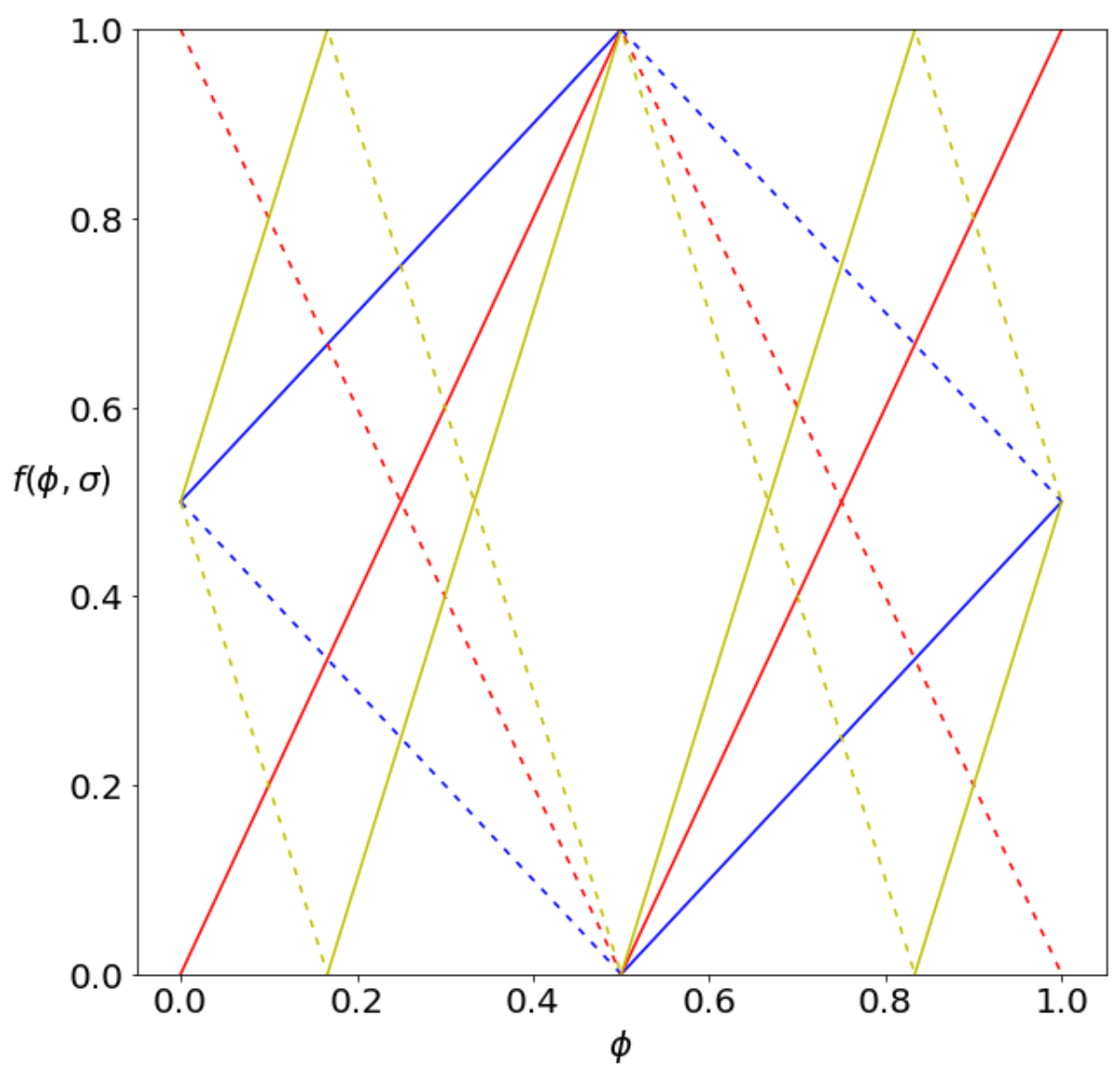}

        \caption{
                \label{fig:samplesetup} 
                Skeleton of the butterfly for $\alpha=\pi/4$.  Solid (dashed) blue lines correspond to $\sigma_{xy}$=1 (-1), solid (dashed) red lines correspond to $\sigma_{xy}$=2 (-2), and solid (dashed) yellow lines correspond to $\sigma_{xy}$=3 (-3).  
        }
\end{figure}
Because our topological map is a combination of the normal butterfly skeleton and the $\pi$-shifted butterfly skeleton we see a doubling of lines associated with odd Hall Conductances, while the even Hall Conductances remain stationary.  

At $\alpha= 0,\pi/4,\pi/2$ the odd numbered Hall band doubling dissapears, and one is left with topological maps associated with Fig. 4.  Notice, however, that $\alpha$=$\pi/2$ is an unphysical region in which $W_0/t \rightarrow \infty$.  Thus we see that there are three topologically distinct phase diagrams associated with the d-density wave problem in an external magnetic field, and that these maps change only at $W_0=0$, and $W_0=4t$.

Using the structure of our obtained skeleton diagrams as a guide we label the Hall conductances for all gaps associated with all butterflies (see Fig.s 7, 8, 9).
\begin{figure}[H]
		 \centering \includegraphics[width=\columnwidth]{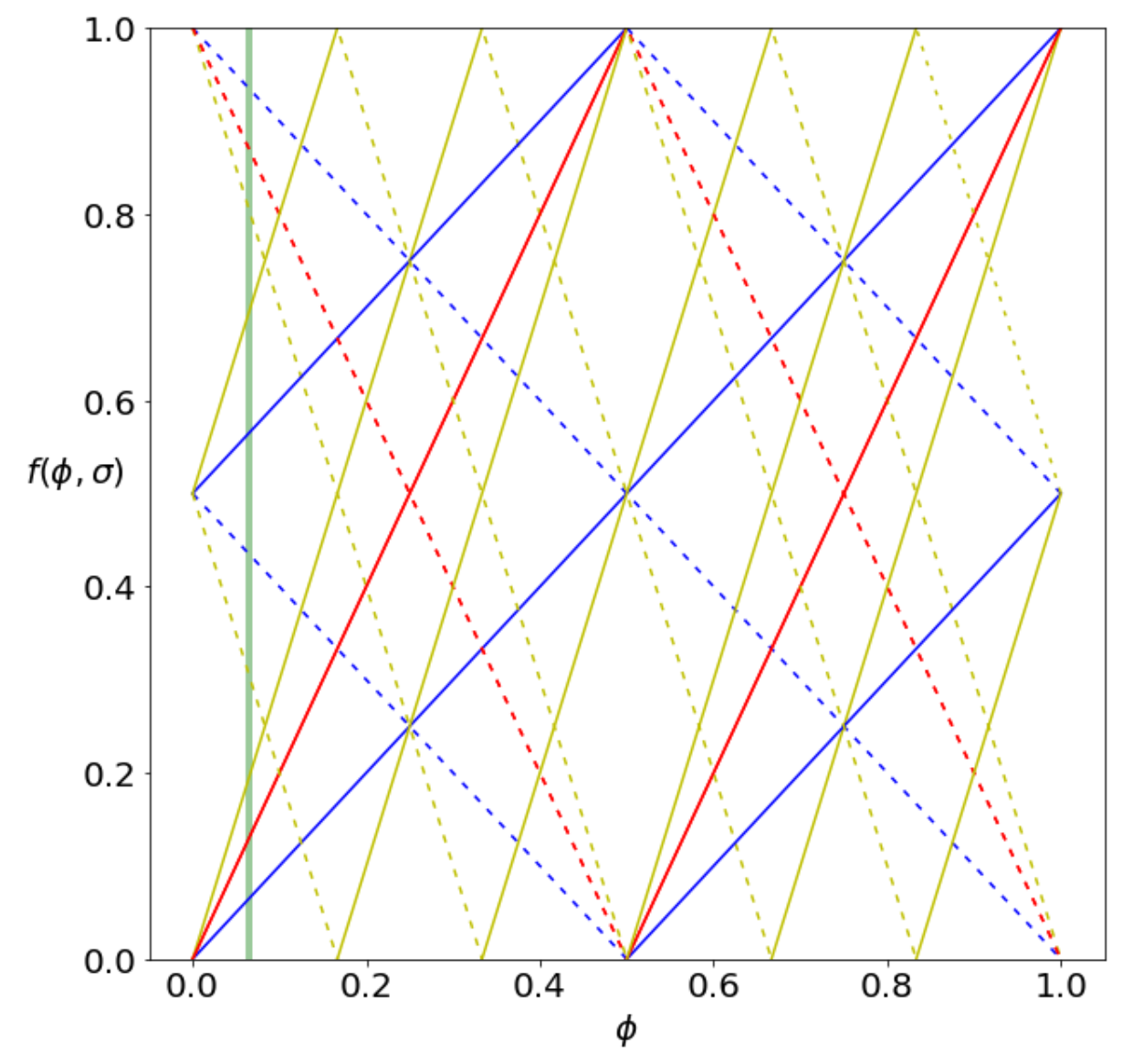}

        \caption{
                \label{fig:samplesetup} 
                Skeleton of the butterflies for $0<\alpha<\pi/4$ .  Solid (dashed) blue lines correspond to $\sigma_{xy}$=1 (-1), solid (dashed) red lines correspond to $\sigma_{xy}$=2 (-2), and solid (dashed) yellow lines correspond to $\sigma_{xy}$=3 (-3).  The vertical green line acts as a guide--indicating that for the regime $0<W_0<4t$, at a fixed flux, one would cross double the amount of odd Chern numbered gaps than those of the typical butterfly as one tunes the Fermi energy from the minimum value of the dispersion's energy to its maximum.
        }
\end{figure}
Due to the odd Hall conductance line doubling for $0<\alpha<\pi/4$ the Hall conductances in low fields near charge neutrality are ``unusual" in the sense that they obey
\begin{equation}
\sigma_{xy} = \pm \frac{e^2}{h} 2(2N+1),
\end{equation}
where $N$ is an integer and we have included a factor of 2 due to spin degeneracy.  The typical integer Quantum Hall conductances persist at the edges of the spectrum near 0 flux and the odd Chern numbered gaps only dissapear completely when $\alpha$=$\pi/4$ where the remaining gaps have
\begin{equation}
\sigma_{xy} = \pm \frac{e^2}{h} 2(2N),
\end{equation}
where, again, we have multiplied by a factor of 2 due to spin degeneracy.
\begin{figure}[H]
		 \centering \includegraphics[width=\columnwidth]{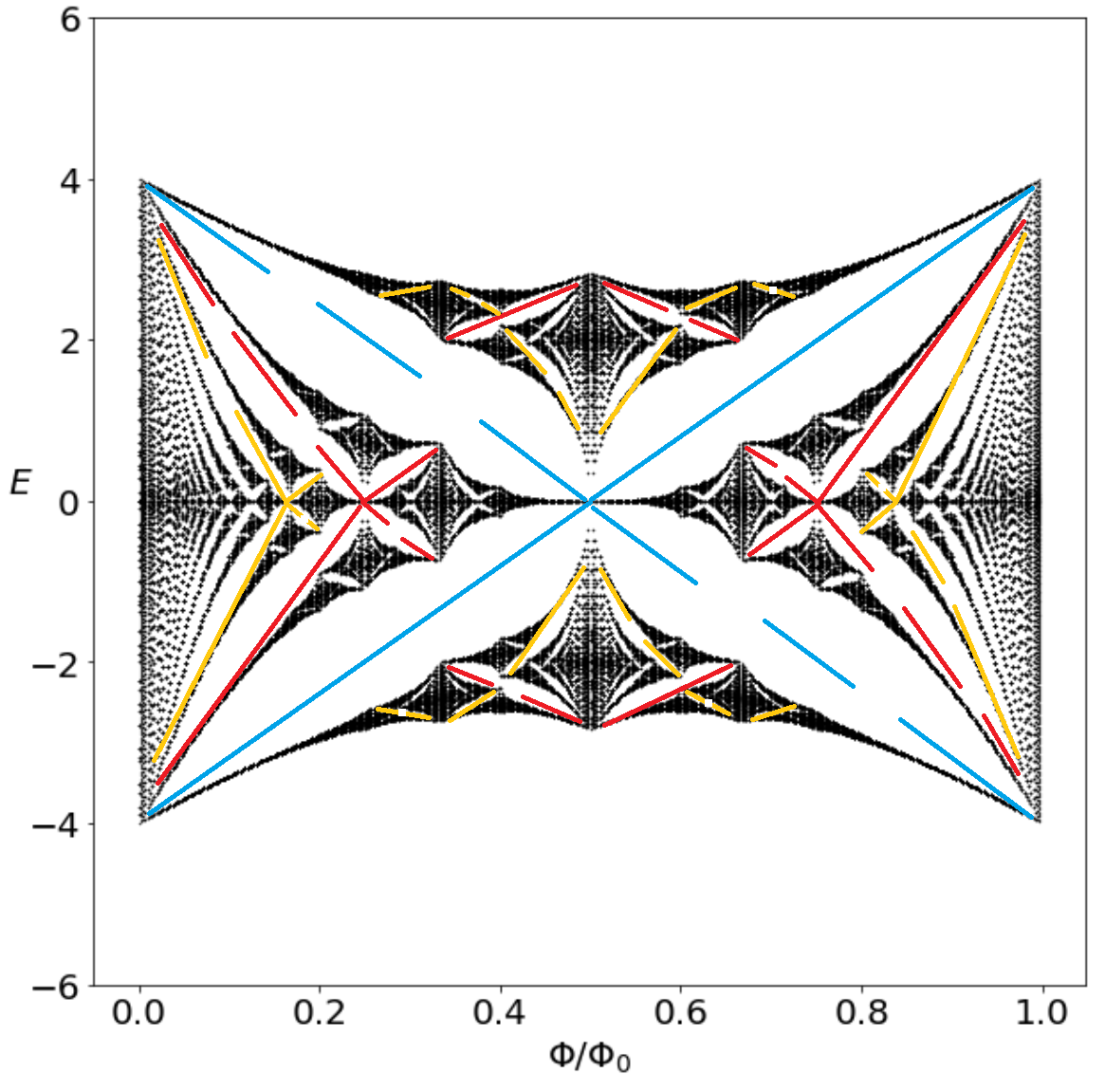}

        \caption{
                \label{fig:samplesetup} 
                Butterfly with labeled characteristic Hall conductances for $\alpha$=0 .  Solid (broken) blue lines correspond to $\sigma_{xy}$=1 (-1), solid (broken) red lines correspond to $\sigma_{xy}$=2 (-2), and solid (broken) yellow lines correspond to $\sigma_{xy}$=3 (-3).
        }
\end{figure}
\begin{figure}[H]
		 \centering \includegraphics[width=\columnwidth]{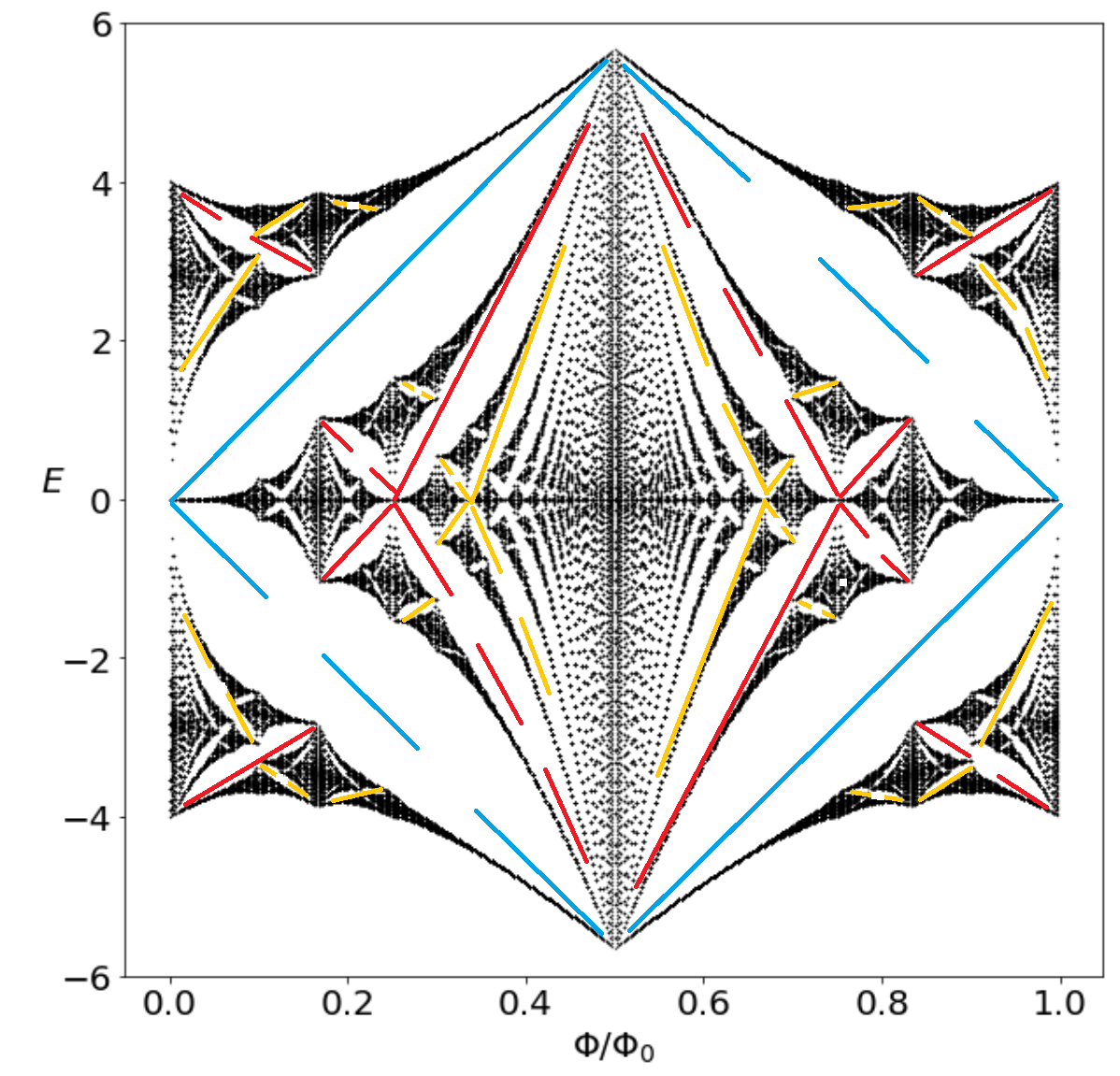}

        \caption{
                \label{fig:samplesetup} 
                Butterfly with labeled characteristic Hall conductances for $\alpha$=$\pi/4$ .  Solid (broken) blue lines correspond to $\sigma_{xy}$=1 (-1), solid (broken) red lines correspond to $\sigma_{xy}$=2 (-2), and solid (broken) yellow lines correspond to $\sigma_{xy}$=3 (-3).
        }
\end{figure}
\begin{figure}[H]
		 \centering \includegraphics[width=\columnwidth]{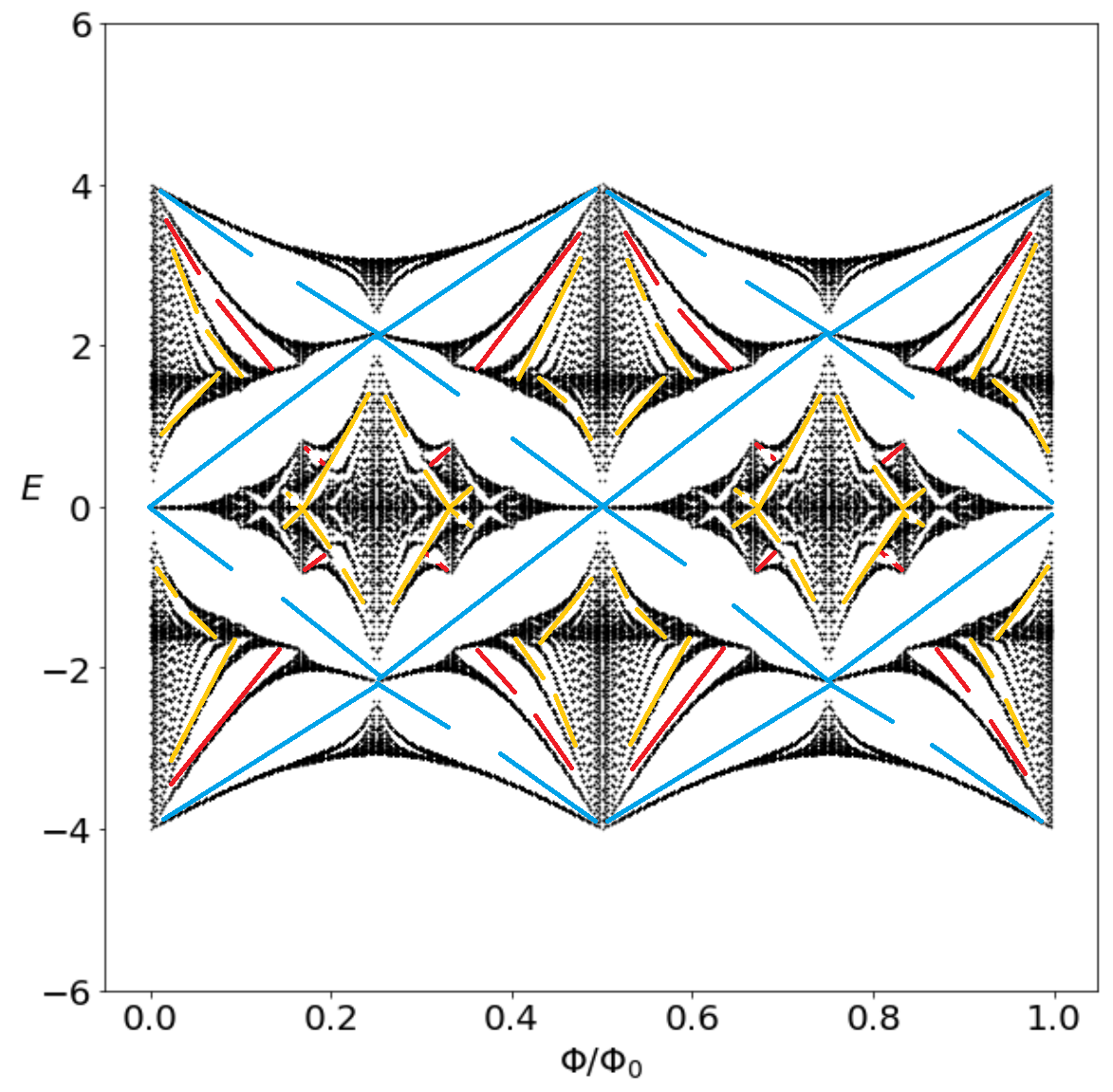}

        \caption{
                \label{fig:samplesetup} 
                Butterfly with labeled characteristic Hall conductances for $\alpha$=$\pi/8$ .  Solid (broken) blue lines correspond to $\sigma_{xy}$=1 (-1), solid (broken) red lines correspond to $\sigma_{xy}$=2 (-2), and solid (broken) yellow lines correspond to $\sigma_{xy}$=3 (-3).
        }
\end{figure}

\section{$p+ip$ Density Wave Order}
 The singlet $\vec{Q}$=$(0,\pi)$ $p_x+i p_y$-density wave state is visualized as both a series of staggered currents pointing along the $x$ direction, and bonds of zero net current that connect nearest neighbors along the $y$ direction\cite{Chetan}.
For this $\vec{Q}$=$(0,\pi)$ $p_x+i p_y$-density wave the Hamiltonian is

\begin{equation} \label{eq:square} 
\begin{split}
H=\sum_{n,m }\Big(-t - i\frac{W_0}{2}(-1)^{n}\Big)e^{i \phi_x}| m+1, n \rangle \langle m, n |\\
+\left(-t + \frac{W_0'}{2}(-1)^{n}\right)e^{i \phi_y}| m, n+1
 \rangle \langle m, n |+ \text{H.C.} \\
\end{split}
\end{equation}
where the density wave order parameter is
\begin{equation}
\langle \psi^{\dagger}(\vec{k}+\vec{Q}) \psi(\vec{k}) \rangle = \pm ( W_0 \text{sin}(k_x)+i W_0' \text{sin}(k_y)).
\end{equation}
In the following we take $W_0=W_0'$, and define $\alpha$ = $\text{arctan}(\frac{W_0}{2t})$.  We plot the butterflies at two characteristic points for a 20$\times$20 lattice in Figs. 10, 11.
\begin{figure}[ht]
		 \centering \includegraphics[width=\columnwidth]{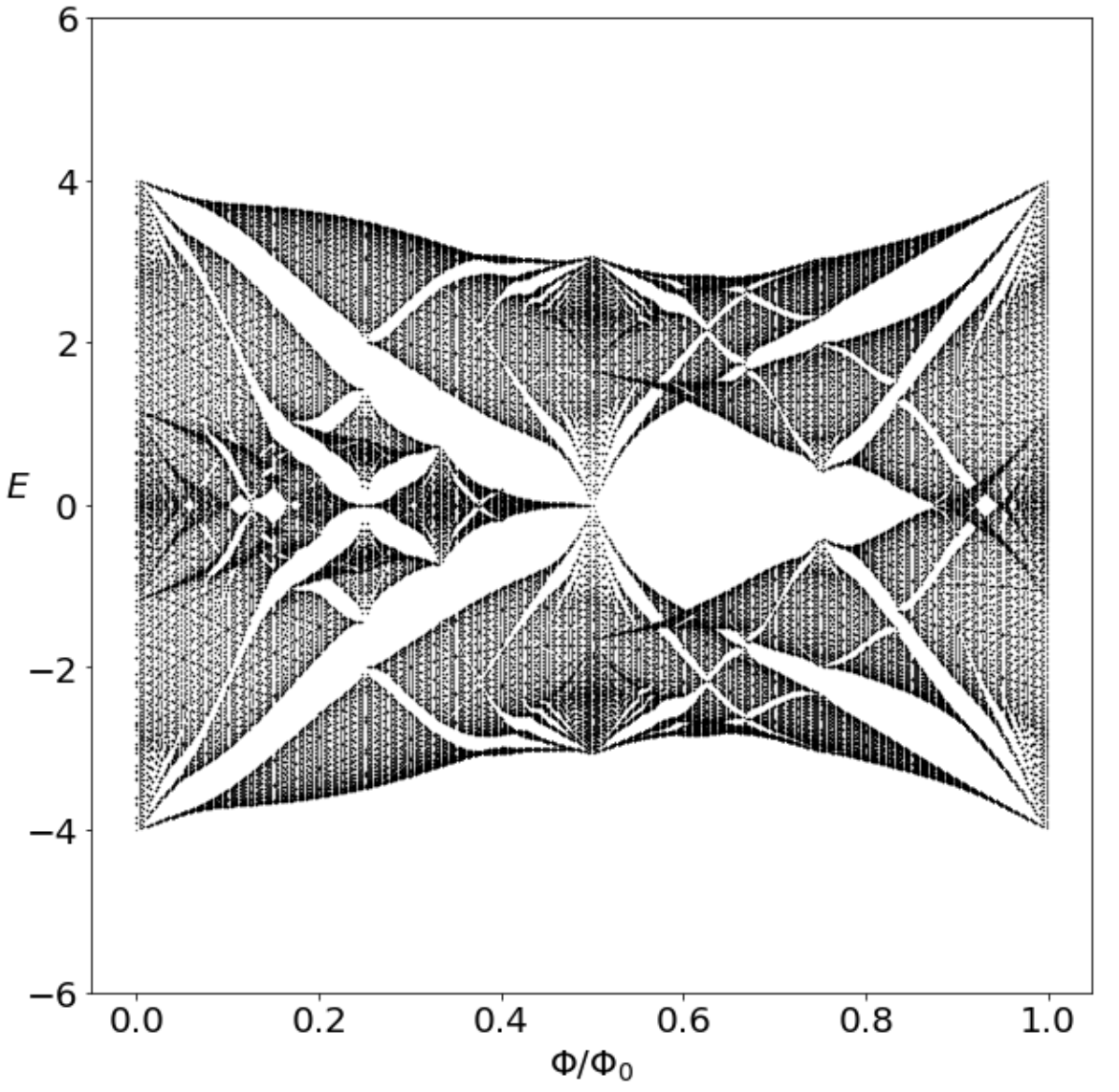}

        \caption{
                \label{fig:samplesetup} 
                Plot of the butterfly for $\alpha=\pi/8$.
        }
\end{figure}
\begin{figure}[ht]
        \centering \includegraphics[width=\columnwidth]{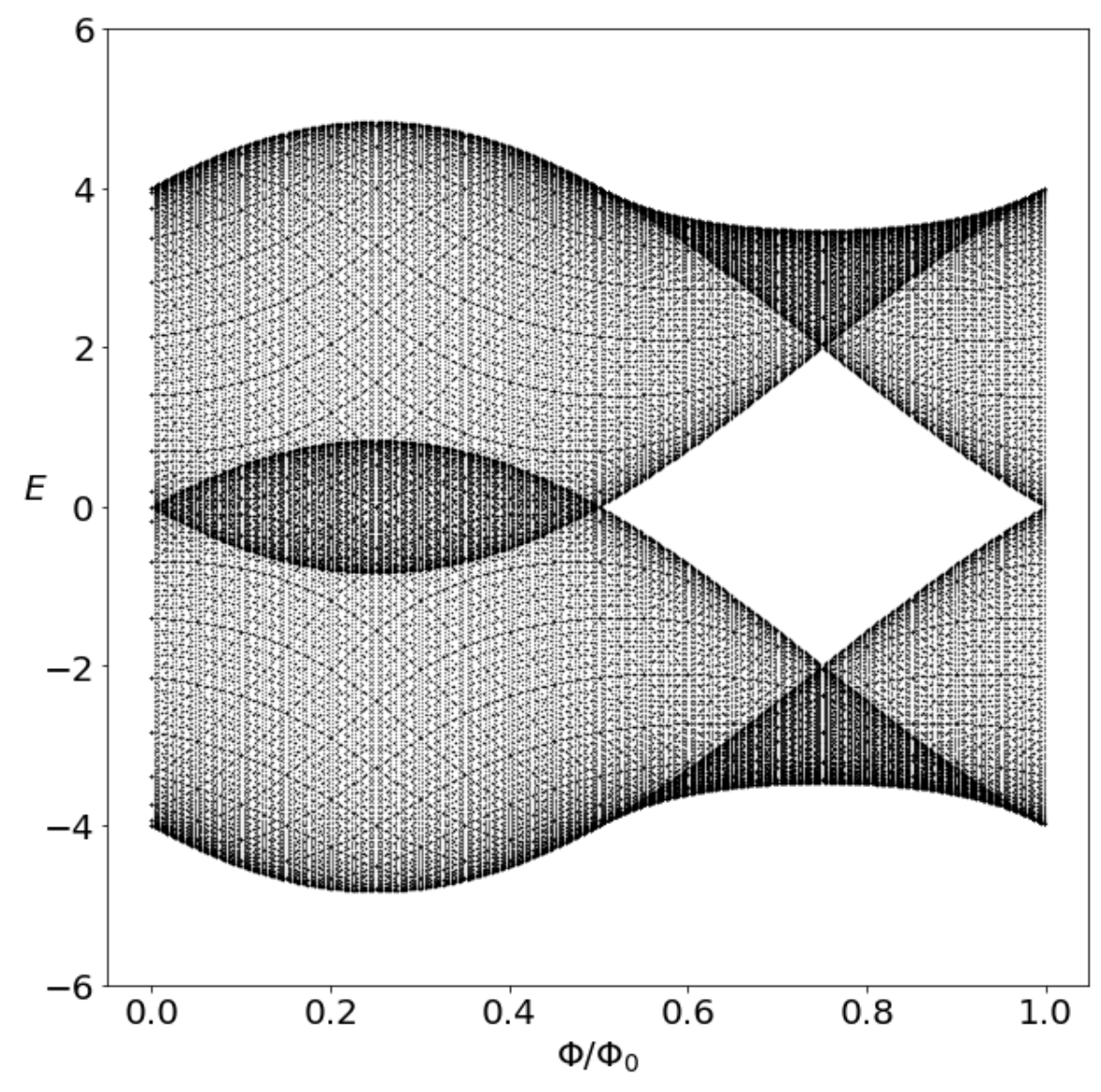}

        \caption{
                \label{fig:samplesetup} 
                Plot of the butterfly for $\alpha=\pi/4$.
        }
\end{figure}
We see that chiral $p$-density wave condensation breaks the butterfly's reflection symmetry about $\pi$ flux, and causes major band gap to collapse--in fact, at $\alpha$=$\pi/4$, when the system is completely dimerized along the $y$ direction and the lattice is composed of disjointed 2$\times L$ ($L$ being the side length of the lattice along the $x$ direction in units of the lattice constant) cylindrical strips of alternating density wave induced fluxes, we find that the Butterfly is completely destroyed and all gaps have collapsed except for a major gap near charge neutrality eminating from $\pi$ flux.  For the $p-ip$-density wave case the spectrum is obtained via a reflection of the $p+ip$ spectrum about $\pi$ flux--implying that the $\sigma_{xy}=0$ gap would be detectable at modest magnetic field strengths.  For either type of chiral $p$-wave condensation at $W_0=2t$ the electrons would not obey any type of Landau quantization of their cyclotron orbits.   Furthermore, because Chern numbers follow a ``zero sum" rule, this major gap at $\alpha=\pi/4$ must have $\sigma_{xy}=0$. 

\section{Multifractal Analysis}
Multifractality is a defining characteristic of wave function fluctuations at criticality\cite{Evers}.  In the following we investigate the nature of the quantum phase transitions that occur as we increase the density wave strength utilizing a basic multifractal analysis of the generalized inverse participation ratio; via this procedure we find that, at a fixed value of magnetic flux, the system undergoes metal-metal transitions seperated by single particle wave functions that exhibit multifractal behavior for a range of density wave strengths.  

The generalized inverse participation ratio (IPR) scales with the system size 
\begin{equation}
P_q = \sum_{m, n} |\psi(r_{m, n})|^{2q} \sim L^{-\tau(q)}
\end{equation}
where the summation is taken over the real space lattice defined by $r_{m, n}$.  The exponents $\tau(q)$, indexed by a continuous variable $q$, are given by $\tau(q)$ = $D_q(q-1)$, where $D_q=d$ for delocalized metallic states and $D_q$=0 for exponentially localized insulating states.  At a critical point the exponents depend on $q$ in a nonlinear fashion--an inherent characteristic of multifractal structures.  In our analyses we focus on the behavior of the system near charge neutrality--thus, to obtain the wave functions pertinent to Eq. 17, we diagonalize the magentic Hamiltonian in real space (as we did when plotting the butterflies) and find the corresponding zero energy eigenvectors of the system for a fixed pair of $\Phi$, and $W_0$.  Numerically there is a difficulty in distinguishing bands from one another when there exists a large degeneracy in the spectrum--thus we choose values of $\Phi/\Phi_0$ = $b/c$ ($b$ and $c$ being coprime integers) such that the degeneracy in the spectrum is minimized for the zero energy eigenvalues.  For this investigation we set $c$=$L^2$, and set $b$ such that both the flux of interest is near typical nontrivial fluxes (such as 1/3, 1/5 etc.), and there exists a zero energy eigenvalue for all $W_0$.  Given this prescription the spectrum will consist of $L^2$ energy eigenvalues with a twofold degenerate zero energy eigenvalue for all values of $W_0$.  For the $d$-density wave case the twofold degeneracy of the zero energy band still hinders our calculation of the IPR as the two bands cannot always be distinguished from one another numerically.  To remedy this we add a small amount of flux $\Delta$=0.000001 to $\Phi/\Phi_0$ which does not alter the spectrum in any appreciable manner but does seperate the twofold degenerate bands from one another enough for us to calculate the IPR of a single band as a function of $W_0$.  This flux offsetting procedure smooths the IPR as a function of $W_0$, but does not alter its global behavior.  We note that for the $p+ip$-density wave case no such offset in flux is needed to distinguish the bands from one another.

As $W_0$ is tuned bands come together and apart, and thus for some energy, flux, and $W_0$, a degeneracy occurs in the spectrum.  At this degeneracy the Chern numbers of the bands participating are no longer well defined, but still follow the requirement that the sum of all Cherns in the spectrum is zero.  For a $26\times26$ lattice we plot both the spectrum and the IPR($q=2$) at $\Phi/\Phi_0=225/676+\Delta$ (see Fig.s 12 and 13).  All listed values of $W_0$ are in units of $t$.
\begin{figure}[ht]
        \centering \includegraphics[width=\columnwidth]{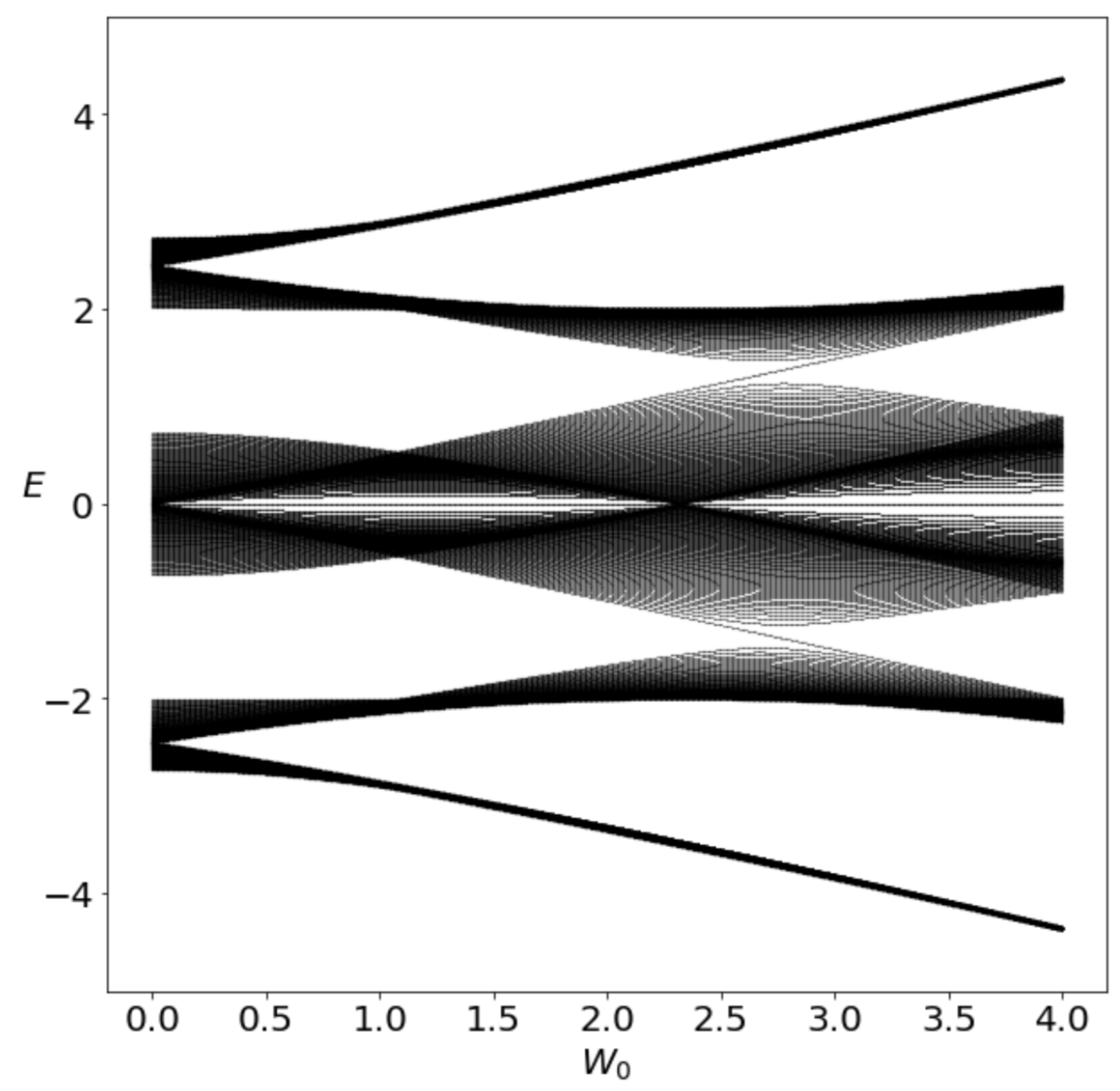}

        \caption{
                \label{fig:samplesetup} 
                Energy versus $d$ density wave strength calculated at $\Phi/\Phi_0=225/676+\Delta$ for a $26\times26$ lattice.
        }
\end{figure}
\begin{figure}[ht]
        \centering \includegraphics[width=\columnwidth]{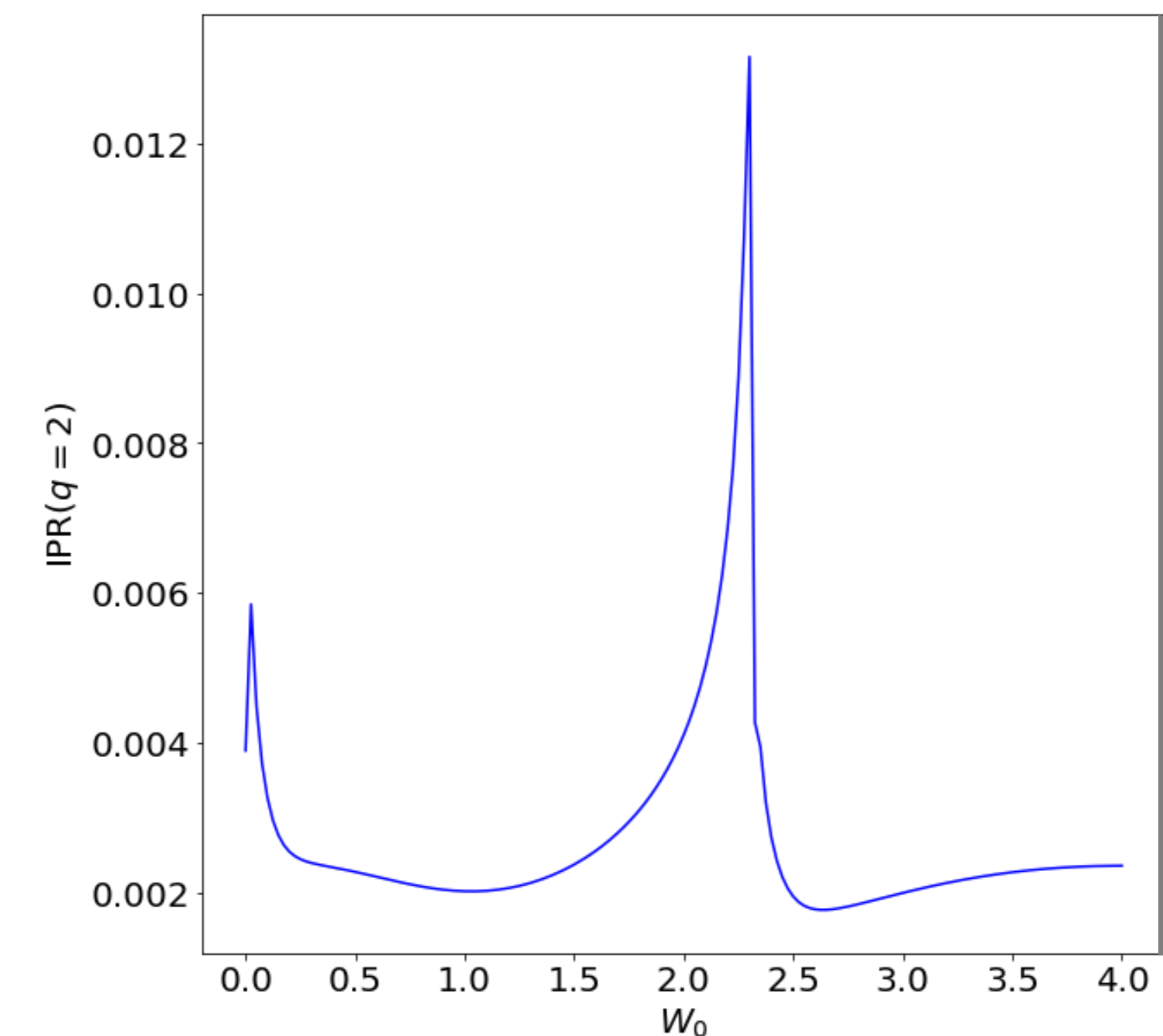}

        \caption{
                \label{fig:samplesetup} 
                Numerically calculated IPR ($q$=2) as a function of $W_0$ of one of the zero energy wave functions for a $26\times26$ lattice at $\Phi/\Phi_0=225/676+\Delta$.  
        }
\end{figure}
At this particular flux there is a band touching event at zero energy occuring at $W_0 \approx 2.3$ which is accompanied by a singular behavior of the IPR$(q=2)$ --indicating a rapid change in the behavior of the single particle wave function fluctuations.  Calculating the multifractal exponents reveals that the zero energy wave functions demonstrate multifractality near this point in parameter space--see Fig. 14.   
\begin{figure}[ht]
        \centering \includegraphics[width=\columnwidth]{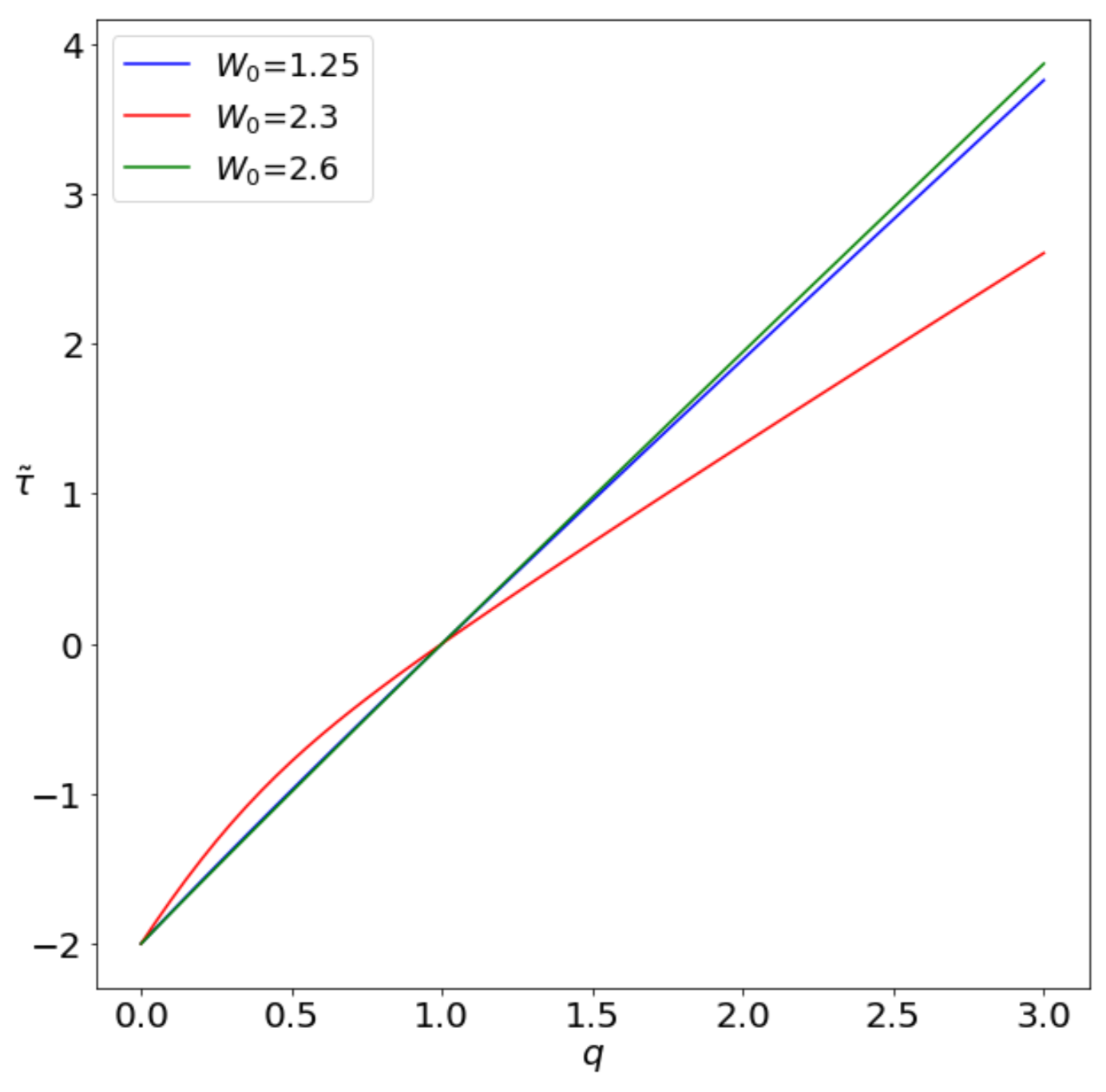}

        \caption{
                \label{fig:samplesetup} 
                Values of -ln$(P_q)$/ln$(L)$ = $\tilde{\tau}(q)$ calculated for a $26\times26$ lattice at $\Phi/\Phi_0=225/676+\Delta$ for three characteristic values of $W_0$.
        }
\end{figure}
At $W_0=2.3$ we find that $\tau(q)$'s leading nonlinear dependance $\approx$ $\frac{-0.517}{q+0.434}$ using a least squares fitting method.  Due to the real space multifractality of the wavefunctions near the central peak shown in Fig. 13, we find that the $d$-density wave controls a metal-metal transition at $\Phi/\Phi_0=225/676+\Delta$ at charge neutrality.  The three distinct regions of phase space follow as approximately: $W_0<$1.5 metallic, 1.5$<W_0<2.5$ critical, and $W_0>$2.5 metallic.  Due to the pseudo-periodic nature of the $d$-density wave term in the Hamiltonian we expect that one more metal-metal transition will occur for much larger values of $W_0$.  In this investigation we have chosen just one of the many band touching events that occur as the $d$-density wave strength is increased, but it should be noted that all such events that we have investigated are effectively identical--thus we claim that all band touching events of this type (near half filling) seperate metallic phases from one another.

For the case of $p+ip$-density wave condensation wavefunctions tend to behave in a localized fashion at $W_0$=2 for all $\Phi>0$ due to the dimerization that occurs in the lattice along the $y$ direction.  We calculate the wavefunctions' multifractal exponents near charge neutrality as we did in the $d$-density wave case and plot the spectrum and IPR with $\Phi$ fixed in Fig.s 15 and 16.
The wavefunctions in this case are multifractal for values near $W_0$=2 (where all bands in the spectrum are connected--similar to the $d$-density wave case), and become ``quasi-localized" (localized to a strip of the lattice) at $W_0$=2 where all gaps in the spectrum are closed. 
\begin{figure}[H]
        \centering \includegraphics[width=\columnwidth]{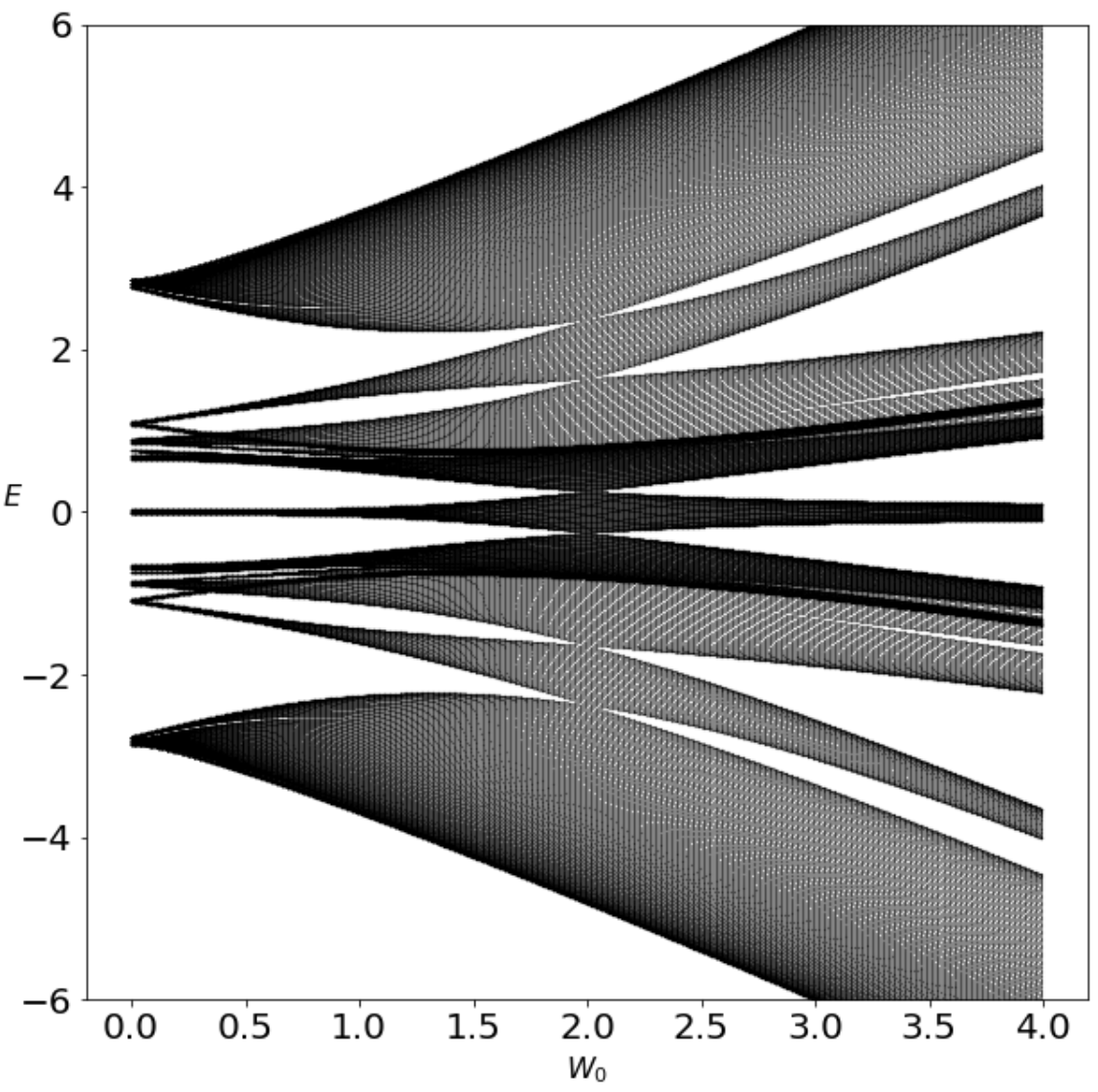}

        \caption{
                \label{fig:samplesetup} 
                Energy versus $p+ip$-density wave strength calculated at $\Phi/\Phi_0=155/676$ for a $26\times26$ lattice.  
        }
\end{figure}
\begin{figure}[ht]
        \centering \includegraphics[width=\columnwidth]{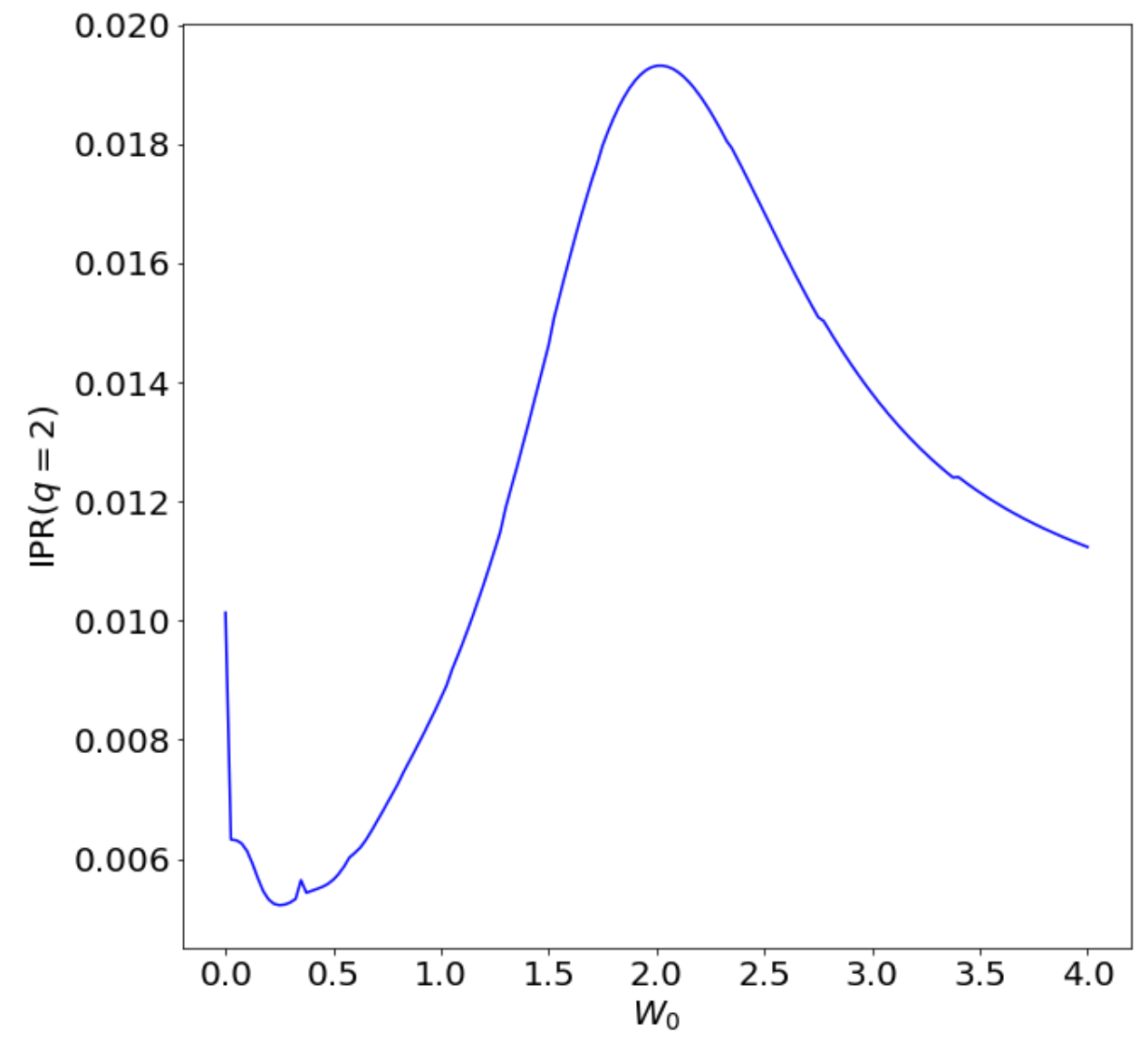}

        \caption{
                \label{fig:samplesetup} 
                Numerically calculated IPR ($q$=2) as a function of $W_0$ of one of the zero energy wave functions for a $26\times26$ lattice at $\Phi/\Phi_0=155/676$. 
        }
\end{figure}
For $W_0$=1.9 we find the nonlinear dependance of $\tau(q)\approx \frac{-0.0832}{q+0.0891}$ using the fitting method mentioned above.  The phase space is seperated into regions given approximately by: $W_0<0.75$ metallic, $0.75<W_0<2$ critical, $W_0=2$ quasi-insulating, 2$<W_0<$3.5 critical, and $W_0>$3.5 metallic.  Just as in the $d$-density wave case, more phase transitions are expected at much larger values of $W_0$.
\begin{figure}[H]
        \centering \includegraphics[width=\columnwidth]{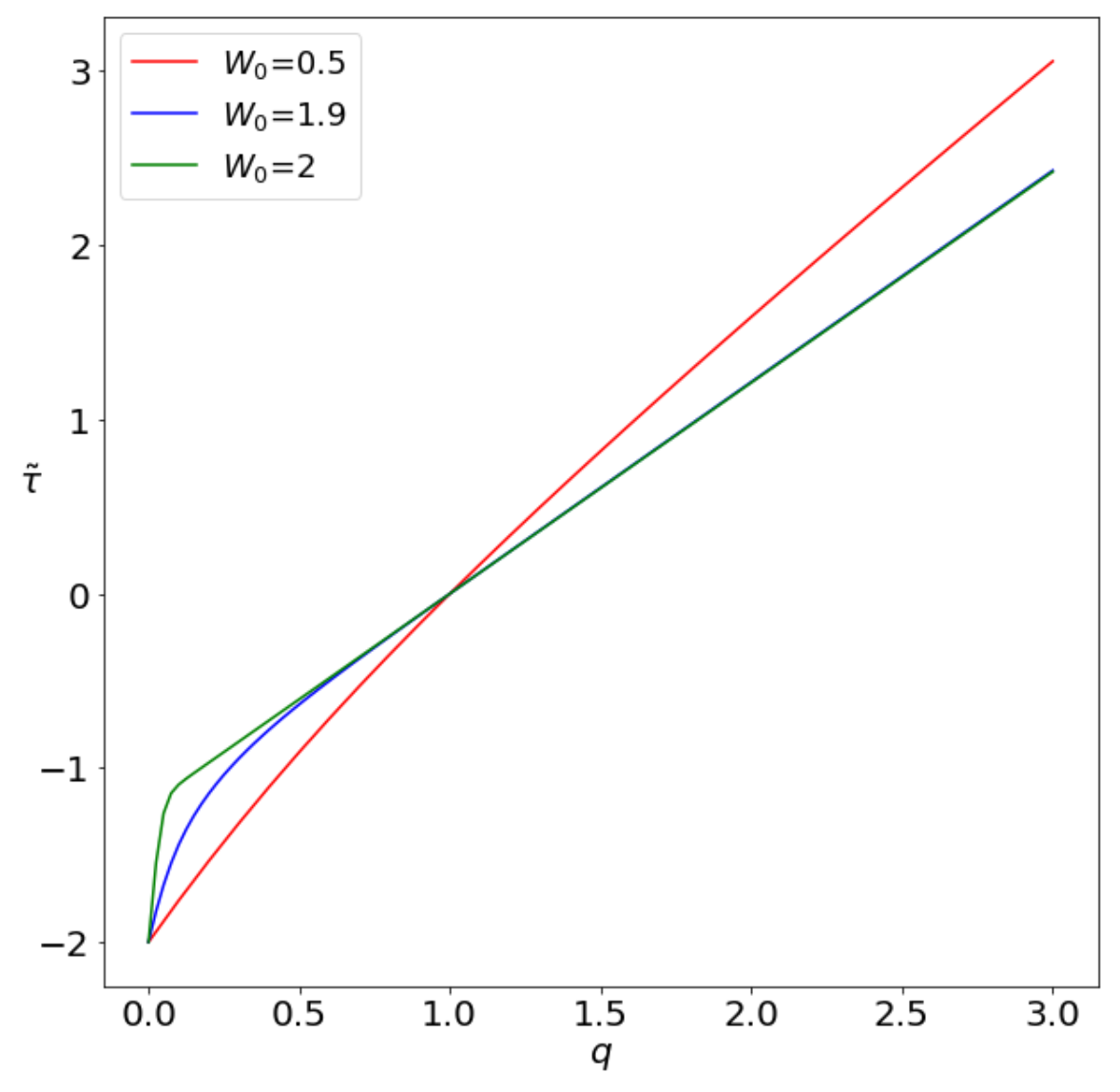}

        \caption{
                \label{fig:samplesetup} 
                 Values of -ln$(P_q)$/ln$(L)$ = $\tilde{\tau}(q)$ calculated for a $26\times26$ lattice at $\Phi/\Phi_0=155/676$ for three characteristic values of $W_0$.
        }
\end{figure}

\section{Discussion}
In this work we have studied and characterized the topologically different forms of the Hofstadter butterflies generated in the presence of density wave condensations in the $d_{x^2-y^2}$ angular momentum channel and investigated the quantum phase transitions that occur at charge neutrality as  density wave strength increases for both the $d$ and $p+ip$ cases.  Directly from the skeleton diagrams obtained for the $d_{x^2-y^2}$-density wave problem we see a doubling in the odd-Hall conductance lines which implies that the density wave strength controls an unusual quantum Hall effect with $\sigma_{xy}$=$\pm\frac{e^2}{h}2(2N+1)$.  Furthermore, we find that the $p+ip$-density wave causes band gap collapses in the butterfly and that, at $W_0/2=t$, the spectrum consists of only a single $\sigma_{xy}$=0 gap.  The effects of density wave states in the presence of an external magnetic field can be detected both at modest magnetic field strengths in 2D square crystal lattices (via a measurement of an unusual quantum Hall effect for the $d$-wave, or via a measurement of the system which shows both a lack of Landau levels and the opening of a $\sigma_{xy}$=0 gap near charge neutrality for the $p-ip$ density wave state), and in optical lattice systems with the appropriate staggered fluxes present. 

Our results have shown that different types of metal-metal transitions, controlled by density wave strength and seperated in phase space by single particle wavefunctions exhibiting multifractality, would be detectable in systems emulating density wave states at non-zero flux at half filling.  These quantum phase transitions occur generically for both density wave condensations near band touching events where at least two distinct bands become connected in the spectrum.  

\begin{acknowledgments}
We would like to thank Steven Durr for helpful discussions. 
\end{acknowledgments}

\appendix
\section{Landau Levels}
To see how the relativistic Landau Levels emerge in the spectrum we expand the tight binding Hamiltonian in the even-odd basis
\begin{equation}
H_0(\vec{k})=-2 \tilde{t}
\begin{bmatrix}
0&e^{-2 i \alpha} \text{cos}(k_x)+\text{cos}(k_y)\\
e^{2 i \alpha} \text{cos}(k_x)+\text{cos}(k_y)&0\\
\end{bmatrix}
\end{equation} 
about one of the charge neutrality points $\vec{k}$=$(\pi/2,\pi/2)$
\begin{equation}
H_0(\vec{k})\approx2 \tilde{t}
\begin{bmatrix}
0&e^{-2 i \alpha} k_x+k_y\\
e^{2 i \alpha} k_x+k_y&0\\
\end{bmatrix}.
\end{equation}
When introducing a magnetic field one makes the substitution 
\begin{equation}
{k_x} \rightarrow {k_x} + \frac{e B {y}}{c} = \tilde{k}_x,
\end{equation}
where $e$ is the electron's charge and $c$ is the speed of light.  Because $\hat{k}_y$ and $\hat{y}$ do not commute with one another we place hats on all crystal momentum and position variables in the Hamiltonian with the understanding that we will work in the real space ( $\hat{k}_y$ = $-i\partial_y$) representation of these operators henceforth.
Rearranging the Schr{\"o}dinger equation 
\begin{equation}
2\tilde{t}
\begin{bmatrix}
0&e^{-2 i \alpha} \hat{\tilde{k}}_x+\hat{k}_y\\
e^{2 i \alpha} \hat{\tilde{k}}_x+\hat{k}_y&0\\
\end{bmatrix}
\begin{bmatrix}
\psi_n\\
\phi_n\\
\end{bmatrix}
= \epsilon_n
\begin{bmatrix}
\psi_n\\
\phi_n\\
\end{bmatrix}
\end{equation} 
yields two decoupled wave equations
\begin{equation}
\epsilon_n^2 \psi_n  = 4 \tilde{t}^2 (e^{-2 i \alpha}\hat{\tilde{k}}_x+\hat{k_y})(e^{2 i \alpha}\hat{\tilde{k}}_x+\hat{k_y}) \psi_n,
\end{equation}
\begin{equation}
\epsilon_n^2 \phi_n  = 4 \tilde{t}^2 (e^{2 i \alpha}\hat{\tilde{k}}_x+\hat{k_y})(e^{-2 i \alpha}\hat{\tilde{k}}_x+\hat{k_y}) \phi_n.
\end{equation}
For the time being we solve Eq. A5.  Foiling out this wave equation we yield

\begin{equation}
\begin{split}
\frac{\epsilon_n^2}{4 \tilde{t}^2} \psi_n(y,k_x) = \Big(\big( \hat{k_x} +  \frac{e B \hat{y}}{c} \big)^2 +\hat{k_y}^2 +\\
\text{cos}(2\alpha)\{\hat{k_x}+ 
\frac{e B \hat{y}}{c} ,\hat{k_y}\} 
-  i \text{sin}(2\alpha)\frac{e B}{c} [\hat{y},k_y]\Big)\psi_n(y,k_x),
\end{split}
\end{equation}
where we have used the fact that $[\hat{k_x},\hat{k_y}]=0$.

We define
\begin{equation}
y_0=k_x \frac{c}{e_0 B}, \; \;
\omega =  \frac{e_0 B}{m c},
\end{equation}
where $e_0$ is the absolute value of the electron charge $e$.  Notice that because the Hamiltonian is independent of $\hat{x}$ we can replace $\hat{k_x}$ with it's eigenvalue $k_x$. 
\begin{figure}[ht]
        \centering \includegraphics[width=.9\columnwidth]{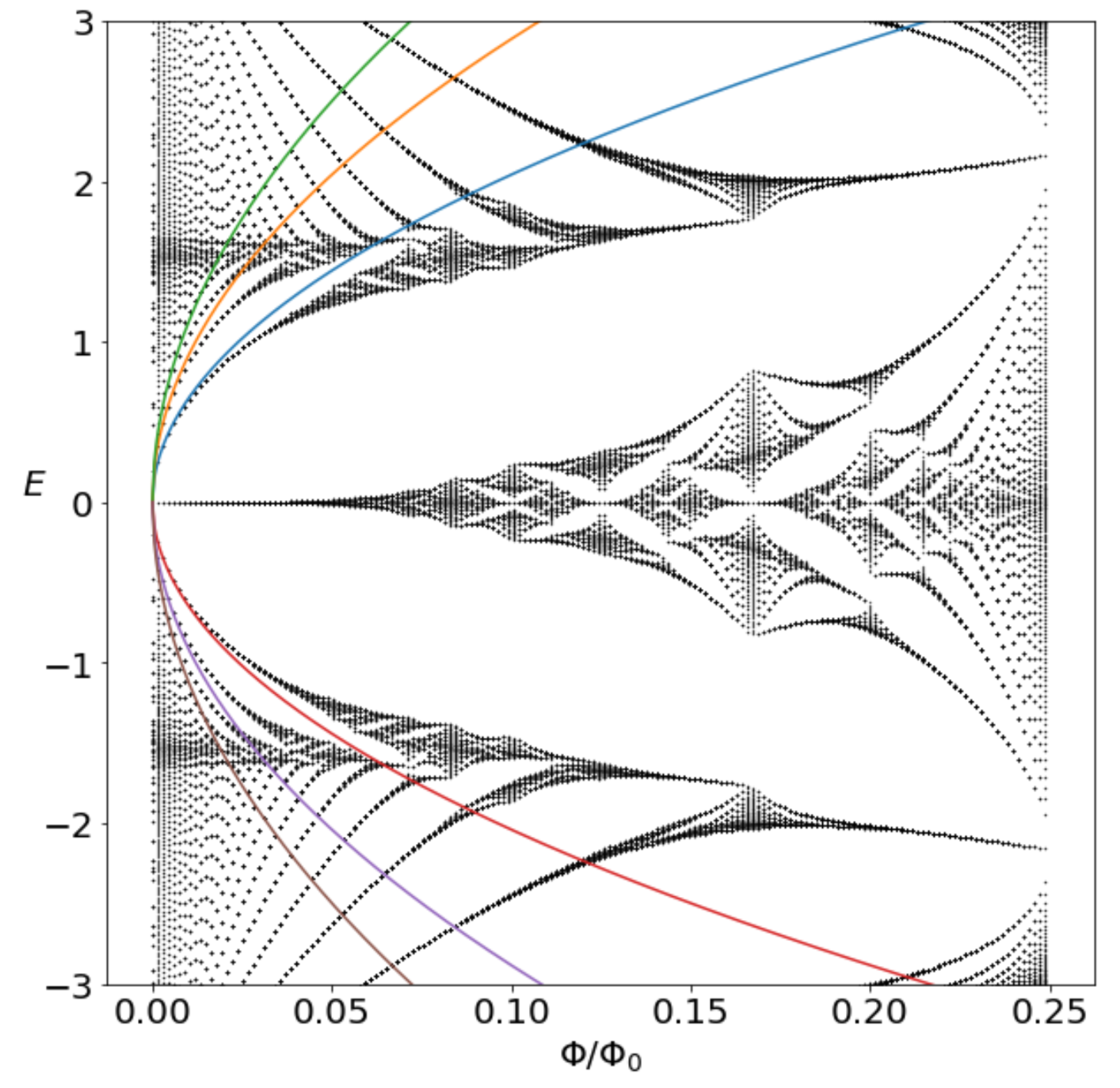}

      \centering  \caption{
                \label{fig:samplesetup} 
                Plot of the butterfly and the associated first few nonzero Landau levels for $\alpha$=$\pi/8$.
        }
\end{figure}
With these definitions in mind we rearrange Eq. A7
\begin{equation}
\begin{split}
\frac{\epsilon_n^2}{8 m \tilde{t}^2} \psi_n(y) =\Big(\frac{1}{2} m \omega^2 ({y}-y_0)^2+\frac{1}{2m}\hat{k_y}^2 - \frac{\omega}{2}\text{sin}(2\alpha)-\\
 \frac{\omega}{2}\text{cos}(2\alpha)\{{y}-y_0,\hat{k_y}\}\Big) \psi_n(y).
\end{split}
\end{equation}
The solutions to this differential equation are of the form
\begin{equation}
\begin{split}
\psi_n(y)= e^{i m \omega \big(\frac{y^2}{2}-y y_0\big)e^{2 i \alpha}}
\Big(C_1 H_n(\sqrt{m \omega |\text{sin}(2 \alpha)|}(y-y_0))+\\
C_2 \; {} _1 F_1(-\frac{n}{2} ;1/2;(m \omega |\text{sin}(2 \alpha)|(y-y_0)^2))\Big),
\end{split}
\end{equation}
where $H_n(y)$ is the Hermite Polynomial of degree $n$ and ${} _1 F_1(-\frac{n}{2};1/2;y^2)$ is the Kummer confluent hypergeometric function. We find the energy eigenvalues of this system by requiring the index of the Hermite polynomials to be of integer value.  Using this prescription we find
\begin{equation} 
\epsilon_n = \pm \tilde{t}\sqrt{8 |\text{sin}(2\alpha)| m \omega n},
\end{equation}
or, in terms of the density wave condensation strength,
\begin{equation} 
\epsilon_n = \pm 2 \sqrt{\frac{e_0 B |W_0| t}{c}n}.
\end{equation}
Solving Eq. A6 in the same fashion yields shifted levels
\begin{equation} 
\epsilon_n = \pm 2 \sqrt{\frac{e_0 B |W_0| t}{c}(n+1)},
\end{equation}
where $n = 0, 1, 2, 3, ...$ for both expressions.  Due to the lack of the zero energy Landau level in Eq. A13 we see that the single particle wave functions will be nonzero only on the even sublattice for index $n=0$, whereas wave functions will have nonzero amplitude on both even and odd sublattices for all $n>0$.  

Solving for the low energy behavior near the $(k_x, k_y)$ = $(-\pi/2,-\pi/2)$ Dirac point yields the same eigenenergy expressions obtained for the ($\pi/2,\pi/$2) case whereas we find the inverse of this even-odd behavior for the $(k_x, k_y)$ = $(\pi/2,-\pi/2)$, $(-\pi/2,\pi/2)$ points.  The Landau level expressions near these points can be obtained by flipping the signs in front of both of the $\text{cos}(2\alpha)$, $\text{sin}(2\alpha)$ terms in Eq. A9.  In this case we find opposite wave function behavior--the single particle wave functions will be nonzero only on the odd sublattice for index $n=0$, and wave functions will have nonzero amplitude on both odd and even sublattices for all $n>0$.  
\begin{figure}[ht]

        \centering \includegraphics[width=.9\columnwidth]{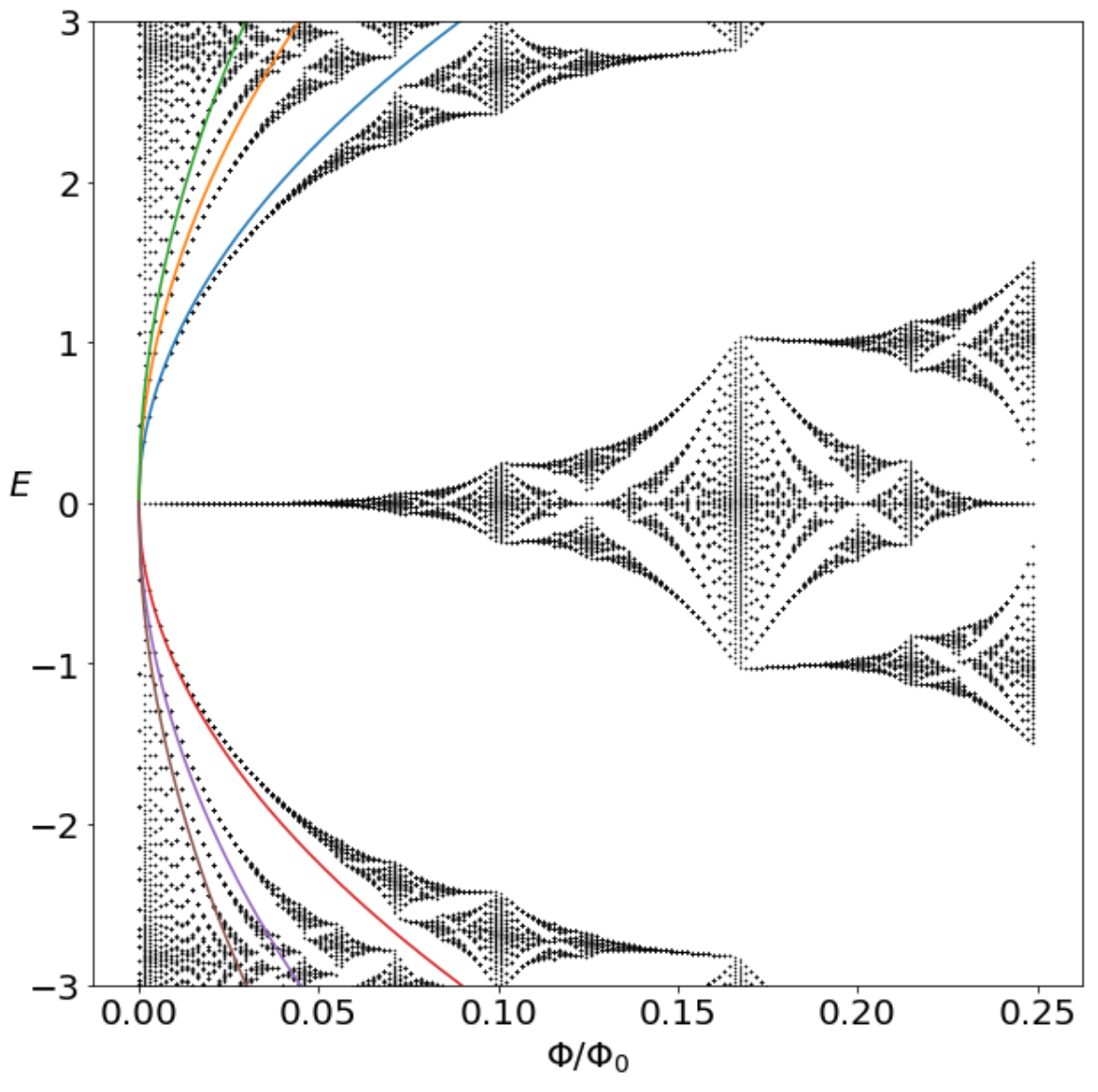}

      \centering  \caption{
                \label{fig:samplesetup} 
                Plot of the butterfly and the associated first few nonzero Landau levels for $\alpha$=$\pi/4$.
        }
\end{figure}

From this analysis we see that for $W_0 \neq 0$ levels emerge from charge neutrality, regardless of the magnitude of $W_0$; this is due to the $d$-density wave's symmetry breaking nature.  We plot characteristic Butterflies and the first few nonzero Landau levels according to Eq.s A12 and A13 in Fig.s 18 and 19.


\begin{thebibliography}{2}
\bibitem{Kim} C. R. Dean {\em et al.} Nature volume {\bf 497}, 598 (2013)
\bibitem{Chetan} 
C. Nayak, Phys. Rev. B \textbf{62},
4880 (2000).
\bibitem{ChakravartyXunGoswami} 
X. Jia, P. Goswami, and S. Chakravarty, Phys. Rev. B \textbf{80},
134503 (2009).
\bibitem{Hofstadter}
D. Hofstadter, Phys. Rev. B \textbf{14},
2239 (1976).
\bibitem{Peierls}
Peierls, R (1933). ``On the theory of diamagnetism of conduction electrons." Z. Phys. \textbf{80} 763-791.
\bibitem{endrodi}
G. Endr\"{o}di, \textit{QCD in magnetic fields: from Hofstadter's butterfly to the phase diagram}, arXiv:1410.8028.
\bibitem{Thouless}
D. J. Thouless, M. Kohmoto, M. P. Nightingale, and M. den Nijs, Phys. Rev. Lett. \textbf{49} (1982), 405.
\bibitem{Naumis}
Naumis, G. Phys. Lett. A \textbf{380} 1772-1780
\bibitem{Evers}
Evers, F., and A. D. Mirlin, Rev. Mod. Phys \textbf{80}, 1355 (2008).
\end{thebibliography}
\end{document}